

\input harvmac

\def\eg{{\it e.g.,}\ }
\def\ie{{\it i.e.,}\ }

\def\p{\partial}

\def\d{\nabla}

\def\({\left (}
\def\){\right )}
\def\[{\left [}
\def\]{\right ]}

%
\Title{\vbox{\baselineskip12pt
\hbox{UCSBTH-95-6; McGill/95-20}
\hbox{gr-qc/9503062}}}
{\vbox{\centerline{ The Value of  Singularities } }}
\baselineskip=12pt
\centerline{
Gary T. Horowitz\footnote{$^1$}{Internet:
gary@cosmic.physics.ucsb.edu}}
 \medskip
\centerline{\sl Department of Physics}
\centerline{\sl University of California}
\centerline{\sl Santa Barbara, CA 93106 USA}
\bigskip
\centerline {and}
\bigskip
\centerline {Robert Myers\footnote{$^2$}{Internet:
rcm@hep.physics.mcgill.ca}}
\medskip
\centerline{\sl Department of Physics}
\centerline{\sl McGill University}
\centerline{\sl Montr\'eal, Qu\'ebec H3A-2T8 Canada}

\bigskip
\centerline{\bf Abstract}
\medskip
We point out that spacetime singularities play a useful role in gravitational
theories by eliminating
unphysical solutions. In particular, we argue that
any modification of general relativity
which is completely nonsingular cannot have a stable ground state.
This argument applies both to classical
extensions of general relativity, and to candidate quantum
theories of gravity.

\Date{}

\def\np {Nucl. Phys. }
\def \pl {Phys. Lett. }

\def \prl {Phys. Rev. Lett. }
\def \pr  {Phys. Rev. }

\def \intj {Int. J. Theor. Phys.}
\def \royp {Proc. Roy. Soc. Lond.}
\gdef \jnl#1, #2, #3, 1#4#5#6{ {\it #1~}{\bf #2} (1#4#5#6) #3}

\lref\unstable{
D.A. Eliezer and R.P. Woodard, \jnl \np, B325, 389, 1989;
\jnl \pr, D40, 465, 1989.}
\lref\field{See for example: N.D. Birrell and P.C.W. Davies,
{it Quantum fields in curved space}, (Cambridge University Press,
1982).}
\lref\sting{See for example: M.B. Green,  J.H. Schwarz and E. Witten,
{\it Superstring Theory}, (Cambridge University Press, 1987).}
\lref\rb{See {\it e.g.,} V. Frolov, M. Markov and V. Mukhanov,
\jnl \pr, D41, 383, 1990; V. Mukhanov and R. Brandenberger, \jnl
\prl, 68, 1969, 1992.}
\lref\pwv{R. Guven, \jnl \pl, B191, 275, 1987;
 G. Horowitz and A. Steif, \jnl \prl, 64, 260, 1990.}
\lref\ks{I. Klebanov and L. Susskind, \jnl \np, B309, 175, 1988.}
\lref\weave{A. Ashtekar, C. Rovelli, and L.Smolin, \jnl \prl, 69, 237, 1992;
 J. Zegwaard, \jnl \pl, B300, 217, 1993.}
\lref\sorkin{L. Bombelli, J. Lee, D. Meyer and R. Sorkin, \jnl \prl, 59,
521, 1987; R. Sorkin, \jnl \intj, 30, 923, 1991.}
\lref\kk{E. Witten, \jnl \np, B195, 481, 1982; D. Brill and H. Pfister,
\jnl \pl, B228, 359, 1989; D. Brill and G. Horowitz, \jnl \pl, B262, 437,
1991.}


General relativity provides an accurate description of a wide range of
gravitational phenomena. However
it is not a complete theory of gravity
since it exhibits spacetime singularities.
This would not be a serious limitation
if singularities were rare, but the theorems
of Hawking and Penrose \ref\hpen{S.W. Hawking and R. Penrose, \jnl \royp, A314,
529, 1970.} show that they are ubiquitous, arising
in large classes of solutions to Einstein's equation.
More precisely, these theorems state that under rather generic conditions,
solutions will be geodesically incomplete.
In many explicit examples, \eg gravitational collapse to form a
black hole, the incompleteness occurs when geodesics terminate in a
region of diverging curvature. Yet while general relativity reaches
an end, physics must continue.
Thus the description provided by general relativity breaks down in a
domain where the curvature is large,
and a proper understanding of such regions
requires new laws of physics.

In a domain of Planck scale curvatures
(\ie $|R_{abcd}|\simeq c^3/(\hbar G)=3.9\times10^{65}{\rm cm}^{-2}$),
the character of gravity will change radically
since its underlying quantum nature will become manifest.
It may be that
the true physics of curvature singularities will not be
revealed until one has fully quantized gravity.
The widespread expectation\foot{A
notable exception is Roger Penrose, who has argued that since the early
universe was very special, the Big Bang singularity must remain in some
form in the ultimate theory.} is certainly that singularities will be
``smoothed out'' or ``resolved'' in the correct theory of quantum gravity.

Alternatively, the physics required to understand curvature singularities
may arise at a classical level. It is quite possible that our classical
description of gravity must be modified before quantization.
For example, classical string theory modifies the equations of motion from
those of general relativity \sting. Typically these modifications
can be understood in the context of a generally
covariant extension of the Einstein action with new higher curvature
interactions (and also higher derivative terms in the metric and any
other matter fields). In the modified equations of
motion, the contribution of these new
terms will be negligible for modest gravitational fields, so these theories
are still consistent with all of the usual experimental tests.
In regions of large curvature, though, the higher curvature terms
can greatly affect the nature of the solutions. In particular,
the contributions from the higher curvature interactions
spoil certain local energy conditions required to prove the
singularity theorems. Hence such theories
evade these theorems, and one might hope
to construct a singularity-free extension of general
relativity \rb.

In this essay, we discuss some general results concerning singularities in
any gravitational theory.
First, we review an argument which shows that
it is impossible to construct a field theory which is completely
free of all curvature singularities.
Second, we argue that on physical grounds,  any reasonable theory will
not ``resolve'' certain classes of timelike singularities. The elimination
of these singularities would lead to a theory without a stable
ground state.
Thus some form of singularity is {\it required} for the theory to be
well-behaved. The latter argument applies to the full quantum theory of gravity
as well as classical extensions of general relativity.
Thus we find that spacetime singularities play
a useful role in gravitational theories, in that they distinguish
certain solutions or states as being unphysical.

To begin,
consider a theory described by a covariant Lagrangian
\eqn\gen{ S = \int d^4x\,\sqrt{-g}\,\left[{R\over16\pi G}+
F(g_{\mu\nu}, \d_\mu, R_{\mu\nu\rho\sigma})\right]}
where $F$ is an arbitrary scalar function of the metric, the curvature
and its derivatives, describing the higher curvature interactions.
The equation for the metric will be the vanishing
of some second rank tensor which is again constructed from the metric,
the Riemann tensor,
its powers and derivatives. We claim that such a theory always has solutions
with unbounded curvature. To see this, consider a gravitational plane wave:
\eqn\pw{ds^2 = -dudv + dx_i dx^i + h_{ij}(u) x^i x^j du^2}
This metric has a covariantly constant null vector $\ell = {\p\,\over\p v}$
and the Riemann tensor is proportional to two powers of $\ell$.
The only nonzero component of the Ricci tensor is
\eqn\ricci{ R_{uu} = -{h^i}_i(u)}
So this metric is a solution to general relativity
(\ie $F=0$) provided $h_{ij}(u)$
is trace-free, with no restriction on its
$u$ dependence. The two independent components of $h_{ij}$ correspond
physically to the amplitudes for the two polarizations of gravitational waves.
The key point is that the metric \pw\ is not only a solution
to general relativity but also to any general theory \gen.
This is because any contraction of $\ell^\mu$
vanishes, so all second rank tensors constructed from the curvature  and its
derivatives must vanish as well \pwv. Since we now have a family of
solutions which
depend on two functions of $u$, we can consider the case where one
(or both) of these
functions diverges. Even though all curvature scalars vanish, one can easily
check that the resulting spacetime has a curvature singularity in the sense
that the gravitational tidal forces are unbounded as the singularity
is approached. This shows that all extensions of general relativity
will have solutions with null singularities.

In a complete physical theory, one may wish to extend the action \gen\
to include covariant matter field contributions. In this case, the above
metric \pw\ with $h^i{}_i(u)=0$ and all matter fields set to zero will still
provide a solution of the general equations of motion. Further, many
of the recent candidates for a theory of quantum
gravity, especially those which attempt to unify gravity
with other interactions, are theories in higher dimensional
spacetimes, and so one may wish to extend these arguments to
an arbitrary dimension. This is trivially accomplished by letting the
indices $i,j$ run over all transverse dimensions.

We now give an argument that all sensible
theories must have timelike singularities
as well. Consider again the general Lagrangian \gen. Since we want this theory
to
reduce to general relativity for long distances and weak curvatures,
the first term (with the fewest derivatives) is simply
the scalar curvature.\foot{We are assuming the cosmological constant vanishes,
although the following argument should extend to nonzero $\Lambda$
(as well as higher dimensions and/or the inclusion of matter fields).}
Now consider the negative mass
Schwarzschild metric. Asymptotically,  the curvature is small, so the
higher order terms in the equation of motion are negligible, and this
metric provides an approximate solution for \gen\ at large $r$.
What happens as we extend the solution in toward $r=0$? Since the field
equations differ significantly from general relativity in regions of large
curvature, it is certainly
possible that the metric remains completely nonsingular. However,
the theory would then have a regular negative energy solution,
and so Minkowski spacetime
would not be stable. In fact, if we want the theory to have any stable lowest
energy solution, it must have singularities in order that one may
discard what would otherwise be pathological solutions. Even if the
theory claims to
have a ground state with $E<0$, we can always start with the Schwarzschild
metric with $M<E$ and argue that it must be singular.
This does not contradict the positive energy theorem
because the higher curvature terms violate the local energy condition required
by this theorem. Thus one has no guarantee
that a ground state exists. In fact we see that removing all
singularities leads to the existence of states with arbitrarily
negative energy.

This simple observation is a powerful constraint on attempts to construct
a singularity-free extension of general relativity. Notice that it is not
necessary to define a singularity in terms of geodesic incompleteness in
order to apply this result. For example, string theory is a modification
of general relativity in which singularities are defined in terms of the
motion of (quantum) test strings. It has been argued that several geodesically
incomplete solutions (\ie singular spacetimes from the viewpoint of
general relativity)
 are nonsingular by this criterion \ref\nonsing{L. Dixon, J. Harvey,
 C. Vafa, and E. Witten, \jnl \np, B261, 678, 1985; M. Rocek and E. Verlinde,
 \jnl \np, B373, 630, 1992.}.
Unless the
string solution which approaches the negative mass Schwarzschild metric
at large distances is singular in the string sense, and hence unphysical,
no stable ground state will exist.

We have been discussing classical modifications of general relativity,
but a similar result must also exist in quantum gravity.
If the classical theory has regular solutions with arbitrarily negative energy,
then it is not likely to lead to a quantum theory with a stable ground state.
Note that these states constitute a new instability beyond those usually
considered in higher derivative theories \unstable.  Also,
unlike the case of a classical charged particle moving in a Coulomb 
potential for which negative energy orbits are confined to a small
region of phase space \unstable,
here we expect a large volume of negative energy solutions since
one can superpose arbitrary numbers of them (at wide separation) with
independent positions and velocities.

Alternatively, suppose the negative mass solutions are classically singular,
but the quantum theory of gravity ``smooths out'' these  singularities. Then
there will again be states of arbitrarily large and negative energy.
In particular, there have been frequent suggestions
that there should be a minimal observable length in quantum gravity
on the order of
the Planck length. The possibility that spacetime is essentially discrete
has been seriously considered in string theory \ks, and the nonperturbative
canonical quantization program initiated by Ashtekar \weave, as well as in
other
approaches \sorkin. If spacetime is fundamentally discrete,
a new mechanism must be found to 
prevent a state which resembles the negative mass Schwarzschild
solution from existing in the theory.

There is one caveat to the above result which should be mentioned.
If a theory has negative energy configurations, one usually cannot
just ``throw them out" since one would expect them to be dynamically produced
in the evolution of certain positive energy states which must be included.
However, it may be that the dynamics prohibits
such production, and so the negative energy states simply decouple from
the theory causing no instability.
An illustration of this is provided by
Kaluza-Klein theory. It is known that if one
allows nontrivial topology, Kaluza-Klein theory admits
nonsingular initial data sets
with arbitrarily negative energy \kk. However if fermions are included,
the theory has two noninteracting
``superselection sectors'', in which the spinors
are periodic or antiperiodic about the compact dimension.
One can show that the energy cannot be negative for the periodic
sector. Thus the pathological configurations
would simply not be a part of the stable and presumably physical sector of
the theory.

However, it is difficult to imagine how such a superselection argument could be
implemented in the case discussed here. By just considering geodesics in the
asymptotic region or the scattering of gravitons in these
configurations, we
know that they do indeed act as localized
negative mass objects which couple to
gravity in the usual way. Hence conventional wisdom would indicate
that they would be created in gravitational interactions, \eg the collision
of gravitational waves.\foot{If all negative mass solutions had nontrivial 
topology, then they could not arise from classical evolution of
initial data on $R^3$. (So Minkowski space would be classically stable.)
However, one would still expect them to arise semiclassically via tunneling
processes that change topology.}
Note that a four dimensional negative mass solution
would destabilize even the periodic sector of a Kaluza-Klein theory.

To summarize, we have argued that rather than being an
undesirable feature
of a theory, singularities play a useful role -- they enable a stable
ground state to exist. Of course
our argument only requires the existence of timelike singularities that
persist for all time. Of greater concern
are the singularities which  form from the evolution of
nonsingular initial conditions. It is possible that
there exists a theory with a stable ground state
in which these
singularities are all removed. However most of the attempts to
eliminate singularities consist of brute force approaches which do not
distinguish between singularities resulting from
collapse and those existing for all time. The lesson we should draw is
that if we wish to find a more complete theory which prohibits
the formation of singularities from regular initial conditions,
we must find a more subtle
mechanism which distinguishes time-independent and time-dependent
strong curvature regions.

\bigskip
We thank Ted Jacobson for useful comments.
RCM would also like to thank the ITP at UCSB for hospitality while this
essay was written. Research at the ITP is supported by NSF Grant No.
PHY94-07194. GTH is supported by NSF Grant No. PHY90-08502, and 
RCM is supported by NSERC of Canada and Fonds FCAR du Qu\'ebec.

\listrefs
\end